\documentclass[onecolumn,nofootinbib,preprintnumbers,longbibliography,showpacs,10pt]{revtex4-1}
\usepackage{array,multirow,graphicx}
\usepackage{dcolumn}
\usepackage{txfonts}
\usepackage{newlfont}
\usepackage{times}
\usepackage{bm}
\usepackage[colorlinks,citecolor=blue,urlcolor=blue,linkcolor=blue]{hyperref}
\usepackage[figtopcap]{subfigure}
\usepackage{color}

\begin{document}
\title{Rotating and non-rotating AdS  black holes in $f({\cal T})$ gravity non-linear electrodynamics}

\author{Salvatore Capozziello}
\email{capozziello@na.infn.it}
\affiliation{Dipartimento di Fisica ``E. Pancini``, Universit\'a di Napoli ``Federico II'',
Complesso Universitario di Monte Sant' Angelo, Edificio G, Via Cinthia, I-80126, Napoli, Italy}
\affiliation{ Istituto Nazionale di Fisica Nucleare (INFN),  Sezione di Napoli,
Complesso Universitario di Monte Sant'Angelo, Edificio G, Via Cinthia, I-80126, Napoli, Italy}
\affiliation{Gran Sasso Science Institute, Viale F. Crispi, 7, I-67100, L'Aquila, Italy}
\affiliation{Laboratory for Theoretical Cosmology, \\Tomsk State University of Control Systems and Radioelectronics (TUSUR), 634050 Tomsk, Russia.}

 \author{Gamal G.L. Nashed}
\email{nashed@bue.edu.eg}
\affiliation {Centre for Theoretical Physics, The British University, P.O. Box
43, El Sherouk City, Cairo 11837, Egypt}
\affiliation {Mathematics Department, Faculty of Science, Ain Shams University, Cairo
11566, Egypt}

\begin{abstract}
 We  derive  new exact charged $d$-dimensional black hole solutions for  quadratic teleparallel equivalent  gravity,  $f({\cal T})=a_0+a_1{\cal T}+a_2{\cal T}^2$, where $\cal T$ is the torsion scalar,  in the case of  non-linear electrodynamics. We give a specific form of electromagnetic   function and find out  the form of the unknown functions that characterize the vielbeins  in presence  of the electromagnetic field. It is possible to show that the black holes behave  asymptotically as AdS solutions   and  contain, in addition to the  monopole and quadrupole terms, other higher order terms whose source is the non-linear electrodynamics field. We calculate the electromagnetic Maxwell field and show that our d-dimensional black hole solutions coincide with the previous   obtained one \cite{2017JHEP...07..136A}.  The  structure of the  solutions show that there is a central singularity that  is much mild in comparison  with  the respective one in General Relativity. Finally,  the  thermodynamical properties of the solutions are investigated  by calculating the entropy, the Hawking temperature, the heat capacity, and other physical quantities.  The most important result of thermodynamics is that the entropy is not proportional to the area of the black hole. This  inanition points out that we must have a constrain   on the  quadrupole term to get  a positive entropy  otherwise we get a negative value.

%\keywords{ $f(T)$ gravitational theories; exact solutions of cubic form of $f(T)$;
%rotating black
%holes; energy.}
\end{abstract}
\date{\today}

\maketitle

%%%%%%%%%%%%%%%%%%%%%%%%%%%%%%%%%%% Section 1 %%%%%%%%%%%%%%%%%%%%%%%%%%%%%%%%%%%%%%%%
\section{Introduction}\label{S1}
%%%%%%%%%%%%%%%%%%%%%%%%%%%%%%%%%%%%%%%%%%%%%%%%%%%%%%%%%%%%%%%%%%%%%%%%%%%%%%%%%%%%%%
Understanding of the gravitational interaction at large scales is considered  a main issue of theoretical physics and cosmology \cite{2016PhRvD..93j4034F}. For example, Einstein's  General Relativity (GR) is not able to explain
  the accelerated expansion epoch of our universe \cite{Riess:1998cb,1999ApJ...517..565P,2013ApJS..208...19H,Eisenstein:2005su,2008PhRvD..78l3532W}. This issue can be solved in the framework of GR if some cosmic flow, having exotic properties, is assumed, the so-called dark energy, or a cosmological constant is involved into the field equation \cite{2003RvMP...75..559P}. Moreover, the rotation curves of spiral galaxy are out the domain of validity of  GR unless one assumes the existence of cold and pressureless dark matter \cite{ 1987gady.book.....B}.

Despite of this state of art, GR has achieved brilliant successes in many aspects like solar system dynamics, gravitational wave detection, relativistic stellar structure up to cosmology \cite{2014arXiv1409.7871W}. Einstein's GR has a question mark when it is confronted with large scales or  when quantization is taken into account. Therefore, there is the necessity for a self-consistent theory that is capable of describing the gravitational interaction ranging from quantum to cosmic scales and coinciding with GR in the  limits where it is successful.

According to this philosophy, there are several proposals to extend or modify GR in view of obtaining a self-consistent theory at any scale.  The so called $f(R)$ gravity is one of these proposal. Here $R$ is the Ricci scalar and, for $f(R)=R$,  its Lagrangian corresponds to the   Hilbert-Einstein Lagrangian and  GR is recovered \cite{2008GReGr..40..357C}.  In some sense, $f(R)$ gravity is the minimal extension of GR. Another proposal  is $f({\cal T})$ gravity  in which ${\cal T}$ represents torsion scalar. Also in this case, for $f({\cal T})={\cal T}$,  the theory reduces to  the so called Teleparallel Equivalent General Relativity (TEGR) which is constructed by the  Weitzenb\" ock geometry and it is   endowed with a nonsymmetric connection,  characterized  by no curvature  and  non-vanishing torsion \cite{2000gr.qc....11087D,Nashed:2018urj,2003gr.qc....12008A,2013AnP...525..339M}. The TEGR torsion tensor   plays the dynamical role of curvature and the vielbein plays the role of metric tensor, that is the gravitational potentials.  Einstein used TEGR theory to construct a unification between  gravitational  and electromagnetism fields
 \cite{2005physics...3046U,Nashed:2003ee,1997PhRvD..56.4689D,2013AnP...525..339M}.   Although, TEGR is constructed using a different geometry from that of GR based on Riemann geometry, that is the   Weitzenb\"{o}ck geometry, TEGR and GR are   completely equivalent from a dynamical point of view. However, assuming generic functions $f(R)$ and $f(\cal T)$,  they are inequivalent \cite{2017PhRvD..95l4024M,2008PhRvD..78l4019F,2009IJMPA..24.1686F}. Therefore, those theories are interesting and can be considered to solve the problems of dark energy and dark matter \cite{2012PhRvD..85l4007C,2011EPJC...71.1752M,Nashed:2016tbj,2011EPJC...71.1797Y,2011PhLB..695..405B,2010JCAP...11..001B,2013RAA....13..757K,2011JCAP...01..009D,
2011CQGra..28u5011C,2011PhRvD..84d3527C,2013PhRvD..88h4042B,2014PhRvD..89h3520C,Nashed:2015pda,2013PhRvD..88j4034N,1996PThPh..95..665S}. In this study, we are considering  $f({\cal T})$ gravitational theories.

There are many applications of $f({\cal T})$ in solar system as well as in cosmological frame. For example, exact solutions, black hole solutions and  stellar models are discussed in \cite{2011PhRvD..84b4042W,2011PhRvD..84h3518F,Nashed:2018qag,Iorio_2015,Nashed:2017fnd,Awad_2018,2012JHEP...07..053G,Awad2019,El_Hanafy_2017,IORIO2016111,Junior_2015,Junior_2015,Junior_2015,Awad_2018,Harko_2014,Capozziello:2012zj,2013JCAP...11..024R,2013GReGr..45.1887N,2010Ap&SS.330..173N,2015JCAP...10..060J,2015EPJC...75...77B}. Spherically symmetric solutions with constant torsion scalar have been derived in \cite{2012ChPhL..29e0402G,2014EL....10510001N}. In the solar system, it is possible to obtain a weak field solution  \cite{2012MNRAS.427.1555I} for the form   $f({\cal T})={\cal T}+\alpha {\cal T}^2$   for which the authors   constrain  the dimensional parameter $\alpha$ \cite{2013MNRAS.433.3584X}. $f({\cal T})$ has extra degrees of freedom which are  related to the non-invariance of the theory under local Lorentz transformations. Recently, an invariant $f({\cal T})$ gravitational theory under local Lorentz transformation has been derived \cite{2012PhRvD..86d4009T}. In the frames of cosmology and spherically symmetric geometry, it is shown that the diagonal ansatz is not a suitable vielbein to be used \cite{2015PhRvD..91j4014R}. A non diagonal spherically symmetric vielbein has been applied to the field equations of $f({\cal T})$ gravity and a weak field solution has been obtained \cite{2015JCAP...08..021I}. Recently, new charged black holes for quadratic and cubic form of $f({\cal T})$ have been derived using flat horizons spacetimes \cite{2017JHEP...07..136A,Nashed:2018cth}. It is the purpose of the present paper to study the effect  of the non-linear electrodynamics in $f({\cal T})$  gravity on a cylindrical  spacetime. This approach could have interesting physical applications both for gravitational and electromagnetic fields.

The layout of the paper is the  following:  In Section \ref{S1} we summarize $f({\cal T})$ gravity
and derive the  field equations in  presence of  non-linear electrodynamics. In Section
\ref{AdsSection} we derive  charged static AdS solutions, analyzing the
 structure  of singularities. In Section \ref{S5} we obtain charged
rotating  AdS solution  in   non-linear electrodynamics in the context of  $f({\cal T})$ gravity.  Section \ref{Thermo} is devoted to   thermodynamics considering  entropy, Hawking temperature, heat capacity, Gibbs free energy. The most interesting feature of these  calculations   is the fact that the entropy is not proportional to the area of the black hole in addition to the possibility of negative values of entropy.    Section \ref{S77} is
devoted to discussion and conclusions.

%We have shown that the entropy of the derived black holes are not proportional to the horizon area and has some regions of the parameter space where the %entropy becomes negative. The main reason for this result is due to the appearance of the parameter $\alpha$ which is related to the higher order correction.

%%%%%%%%%%%%%%%%%%%%%%%%%%%%%%%%%%% Section 1 %%%%%%%%%%%%%%%%%%%%%%%%%%%%%%%%%%%%%%%%
\section{Basic concepts  of $f({\cal T})$ gravity} \label{S1}
In  Riemannian geometry, the metric of the spacetime has the form
\begin{equation} \label{met}
ds^2:=g_{\mu \nu} dx^\mu dx^\nu,\end{equation} where  $g_{\mu \nu}$ is a second order symmetric tensor. Using the vielbein one can write Eq. (\ref{met}) as
\begin{equation} \label{met1}
ds^2:=g_{\mu \nu} dx^\mu dx^\nu=\eta_{i j}\vartheta^i \vartheta^j, \quad \textrm{where} \quad \vartheta^i={h^i}_\mu dx^\mu, \end{equation} with $\eta_{i j}$ being the  Minkowskian metric that is defined as:
$\eta_{i j}=diag(-1,+1,+1,\cdots,+1)$ and ${h^i}_\mu$ is the covariant vielbein that satisfies the orthogonality conditions
\begin{equation} {h^i}_\mu{h_j}^\mu={\delta_j}^i, \qquad \qquad  {h^i}_\mu{h_i}^\nu={\delta_\mu}^\nu. \label{orth}
\end{equation}
To construct a spacetime with vanishing curvature and a non-vanishing torsion one  has to define  the  Weitzenb\"ock connection  which is
\begin{equation}\label{con}  {\Gamma^\lambda}_{\nu \mu}= {h_i}^\lambda~ \partial_\mu h^{i}{_{\nu}}=- h^{i}{_{\nu}} \partial_\mu {h_i}^\lambda, \qquad \textrm{where}  \qquad \partial_\mu=\frac{ \partial}{ \partial x^\mu}.
\end{equation}
Using Eq. (\ref{con}), the  torsion and contorsion tensors are
\begin{eqnarray} \label{q4}
  {T^\alpha}_{\mu \nu}  :=
{\Gamma^\alpha}_{\nu \mu}-{\Gamma^\alpha}_{\mu \nu}, \qquad \qquad
{\gamma^{\mu \nu}}_\alpha  :=
-\frac{1}{2}\left({T^{\mu \nu}}_\alpha-{T^{\nu
\mu}}_\alpha-{T_\alpha}^{\mu \nu}\right).
\end{eqnarray}
From Eqs. (\ref{q4}), one can define the superpotential tensor
\begin{equation}\label{q5}  {S^{\lambda \mu}}_{\nu}=\frac{1}{2}\left({\gamma^{\mu\nu}}_\alpha+\delta^\mu_\alpha{T^{\beta
\nu}}_\beta-\delta^\nu_\alpha{T^{\beta \mu}}_\beta\right).
\end{equation} From Eqs. (\ref{q4}) and (\ref{q5}) the torsion scalar is provided as
\begin{equation}\label{Tor_sc}
{\cal T}:= {T^\alpha}_{\mu \nu} {S_\alpha}^{\mu \nu}.
\end{equation}
The Lagrangian of TEGR theory is constructed from the torsion scalar given by Eq. (\ref{Tor_sc}).

Let us now  consider  $f({\cal T})$  gravity minimally coupled with  non-linear electrodynamics. Thence, the action of this theory is given by
\begin{equation} \label{act} {\cal L}:=\frac{1}{2\kappa}\int |h|f({\cal T})~d^{d}x+\int |h|{\cal L({\cal F})}~d^{d}x,\end{equation}
 with $|h|=\sqrt{-g}=\det\left({h^a}_\mu\right)$ being the determinant of the metric and
$\kappa$    a dimensional constant with the form $\kappa =2(d-3)\Omega_{d-1} G_d$, with
$G_d$ being  the Newtonian
gravitational  constant in $d$-dimensions and
$\Omega_{d-1}$   a   $(d-1)$-dimensional unitary volume whit the form
\begin{equation}
 \Omega_{d-1} = \frac{2\pi^{(d-1)/2}}{\Gamma((d-1)/2)},
 \end{equation}
 where $\Gamma$  is the $\Gamma$-function (when   $d = 4$, it is
$2(d-3)\Omega_{d1} = 8 \pi$).  The electromagnetic Lagrangian   ${\cal L({\cal F})}$ is  gauge-invariant   and depends on the  invariant ${\cal F}$  defined as ${\cal F} = \frac{1}{4}{\cal F}_{\alpha \beta}{\cal F}^{\alpha \beta}$ \cite{plebanski1970lectures}. The  antisymmetric Faraday tensor is defined as
 \begin{equation} {\cal F}_{\alpha \beta}= {\cal E}_{\alpha, \beta}- {\cal E }_{\beta, \alpha},\end{equation}
 where ${\cal E}_\mu$ is its gauge potential 1-form. In the Maxwell theory,  the    Lagrangian ${\cal L({\cal F})}$  is ${\cal L({\cal F})}=4\cal F$. Here, we  consider a more general choice of the electromagnetic Lagrangian. From Action  (\ref{act}), the non-linear electrodynamics is described by  nonlinear terms in  ${\cal F}_{\alpha \beta}$ and its invariants. However, we can provide a dual  representation in terms of an auxiliary field ${\cal P}_{\alpha \beta}$ . This method is
proved to be highly benefit to derive  exact solutions in GR,  specifically for the electric case \cite{1999PhLB..464...25A,1987JMP....28.2171S}. The dual form can be obtained adopting the  Legendre transformation below:
\begin{equation} \label{Ha}
{\cal \aleph}=2{\cal F} {\cal L}_{\cal F}-{\cal L}, \quad \textrm{where} \qquad {\cal L}_{\cal F}=\frac{\partial {\cal L}}{\partial {\cal F}},
\end{equation}
 where  ${\cal \aleph}$ is an arbitrary function  depending on the invariant ${\cal P}$,  defined as ${\cal P}=\frac{1}{4}{\cal P}_{\alpha \beta}{\cal P}^{\alpha \beta}$. From Eq. (\ref{Ha}), non-linear electrodynamics can be recast in terms of ${\cal P}$  according to the formulas 
\begin{equation} \label{rel}   {\cal P}_{\mu \nu}={\cal L}_{\cal F} {\cal F}_{\mu \nu}, \qquad \qquad  {\cal F}_{\mu \nu}={\cal \aleph}_{\cal P} {\cal P}_{\mu \nu}, \qquad \qquad {\cal L}=2{\cal P} {\cal \aleph}_{\cal P}-{\cal \aleph},\end{equation}
 where the standard Maxwell theory is  obtained for ${\cal L}_F=1$. As it is clear from the above equations, ${\cal \aleph}$ is a function of ${\cal P}$, where \cite{1999PhLB..464...25A,1987JMP....28.2171S} $${\cal \aleph}_{\cal P}=\frac{\partial {\cal \aleph}}{\partial {\cal P}}.$$
The variation of Lagrangian  (\ref{act}) with respect to the vielbeins  leads to
\begin{eqnarray}\label{q8a}
& &\zeta^\nu{}_\mu={S_\mu}^{\rho \nu} \partial_{\rho} {\cal T}
f_{\cal TT}+\left[h^{-1}{h^i}_\mu\partial_\rho\left(h{h_i}^\alpha
{S_\alpha}^{\rho \nu}\right)-{T^\alpha}_{\lambda \mu}{S_\alpha}^{\nu \lambda}\right]f_{\cal T}
-\frac{f}{4}\delta^\nu_\mu +\frac{1}{2}\kappa{{{\mathfrak{
T}}^{{}^{{}^{^{}{\!\!\!\!\scriptstyle{nlem}}}}}}}^\nu_\mu \equiv0,
\end{eqnarray}
and the Maxwell  equations for non-linear electrodynamics become \cite{1999PhLB..464...25A}
\begin{equation} \label{maxf}
\partial_\nu \left( \sqrt{-g} {\cal P}^{\mu \nu} \right)=0.\end{equation}
The stress-energy tensor of  non-linear electrodynamic is 
\begin{equation} \label{max1}
{{{\mathfrak{
T}}^{{}^{{}^{^{}{\!\!\!\!\scriptstyle{nlem}}}}}}}^\nu_\mu:=2({\cal \aleph}_{\cal P}{\cal P}_{\mu \alpha}{\cal P}^{\nu \alpha}-\delta_\mu^\nu [2{\cal P}{\cal \aleph}_{\cal P}-{\cal \aleph}]). \end{equation}
It is worth saying  that Eq. (\ref{max1}) has  a non-vanishing  trace unlike the stress-energy tensor coincides with the Maxwell one. It is worth  noticing that the electric field of   linear electrodynamics is obtained as
\begin{eqnarray} \label{Max3}
{\cal E} = {\cal F}_{tr} ={\cal \aleph}_{\cal P}{\cal P}_{tr}.\end{eqnarray}
In this context, black hole solutions can be found.

\section{Anti-de-Sitter  black hole solutions in non-linear electrodynamics}
\label{AdsSection}

Let us search now for   charged AdS black hole solutions in non-linear electrodynamics assuming, in general,   $d$-dimensions in the framework  of $f({\cal T})$ gravity. Using the following vierbein diagonal ansatz   in $d$-dimensions ($t$, $r$,
$\eta_1$, $\eta_2$, $\cdots$, $\eta_{n}$, $\xi_1$, $\xi_2$ $\cdots$ $\xi_l$), with  $l=1,2 \cdots$ $d-n-2$, in
which   $0\leq r< \infty$, $-\infty < t < \infty$, $0\leq \eta_n< 2\pi$ and $-\infty <\xi_k < \infty$, we
assume the vielbein \cite{Capozziello:2012zj,Nashed:2018cth}:
\begin{equation}\label{tetrad}
\hspace{-0.3cm}\begin{tabular}{l}
   $\left({h^{i}}_{\mu}\right)=\left( \sqrt{A(r)}, \; \frac{1}{\sqrt{A(r)g(r)}}, \; r, \;
r, \; r\;\cdots \right)$,
\end{tabular}
\end{equation}
which corresponds to the metric
\begin{equation}
\label{m2}
ds^2=
-A(r)dt^2+\frac{1}{A(r)g(r)}dr^2+r^2\left(\sum_{i=1}^{n}d\eta^2_i+\sum_{l=1}^{d-n-2}
d\xi_l^2\right),
\end{equation}
where    $A(r)$ and $g(r)$ are functions depending only on the radial
coordinate $r$. Substituting the vielbein  (\ref{tetrad}) into the torsion scalar in (\ref{Tor_sc}), we get
\begin{equation}\label{df}
{\cal T}=(d-2)\frac{A'g}{r}+(d-2)(d-3)\frac{Ag}{r^2},
\end{equation}
where $A'(r)\equiv\frac{dA(r)}{dr}$ and $g'(r)\equiv \displaystyle\frac{dg(r)}{dr}$. Finally, since the
$f({\cal T})$ power law gravity seems the model with the best agreement with  observational data
  \cite{2013PhRvD..88j3010N,2016JCAP...08..011N,2018JCAP...08..008B},  we will focus on 
 the choice
\begin{equation}\label{powellaw}
 f({\cal T})=a_0+a_1{\cal T}+a_2{\cal T}^2,
\end{equation}
where $a_0$,  $a_1$ and $a_2$ are the model parameters.

\subsection{Asymptotically static  AdS black holes}\label{S2}

Inserting the vielbein (\ref{tetrad}) into   field Eqs. (\ref{q8a}) and (\ref{maxf}),  we
obtain the following non-vanishing components:
%\newpage
\begin{eqnarray}\label{df1}
& & \zeta^r{}_r= 2{\cal T}f_{\cal T}+2a_0-f-4\aleph=0,\nonumber\\
& &   \zeta^{\eta_1}{}_{\eta_1}= \zeta^{\eta_2}{}_{\eta_2}=\cdots \cdots
=\zeta^{\eta_{n}}{}_{\eta_{n}}=\zeta^{\xi_1}{}_{\xi_1}= \zeta^{\xi_2}{}_{\xi_2}=\cdots \cdots
=\zeta^{\xi_{d-n-2}}{}_{zeta_{d-n-2}}\nonumber\\
& &=   \frac{f_{{\cal TT}}
[r^2{\cal T}+(d-3)A]{\cal T}'}{r}+\frac{f_{\cal T}}{2r^2}\Biggl\{2r^2gA''+rg'(2(d-3)A+rA')+g(2(d-3)^2A+(3d-8)rA')\Biggr\}-f+2a_0-4\aleph\nonumber\\
& &+\frac{8q'g\aleph'}{q'g'+2gq''}=0, \nonumber\\
& & \zeta^t{}_t=\frac{2(d-2)Agf_{{\cal TT}}
{\cal T}'}{r}+\frac{(d-2)f_{\cal T}}{r^2}\Biggl\{2[(d-3)Ag+rgA']+rAg'\Biggr\}
-f-4\aleph+2a_0=0,\nonumber\\
& &
\end{eqnarray}
where $q(r)\equiv q$ is the gauge 1-form of the non-linear electrodynamics that is defined as ${\cal P}_{tr}=q'$ and $q'=\frac{dq}{dr}$,  $q''=\frac{d^2q}{dr^2}$.  In the case  of  $f({\cal T})$ with the form  given by (\ref{powellaw}), the above equations reduce to
\begin{eqnarray}
\label{df31}
& & \zeta^r{}_r=  a_1{\cal T}+3a_2{\cal T}^2+a_0-4\aleph=0,\\
\label{df32}
& & \zeta^{\eta_1}{}_{\eta_1}= \zeta^{\eta_2}{}_{\eta_2}=\cdots \cdots
=\zeta^{\eta_{n}}{}_{\eta_{n}}=\zeta^{\xi_1}{}_{\xi_1}= \zeta^{\xi_2}{}_{\xi_2}=\cdots \cdots
=\zeta^{\xi_{d-n-2}}{}_{\xi_{d-n-2}}\nonumber\\
  &&=\frac{2a_2[r^2{\cal T}+(d-3)A]{\cal T}'}{r}+\frac{(a_1+2a_2 {\cal T})}{2r^2}\Biggl\{2r^2gA''+rg'(2(d-3)A+rA')+g(2(d-3)^2A+(3d-8)rA')\Biggr\}\nonumber\\
& &-a_1{\cal T}-a_2{\cal T}^2+a_0-4\aleph+\frac{8q'g\aleph'}{q'g'+2gq''}=0,\\
\label{df33}
& &\zeta^t{}_t= \frac{4a_2(d-2)Ag{\cal T}'}{r^2}+\frac{(a_1+2a_2 {\cal T})(d-2)}{r}\Biggl\{2[(d-3)Ag+rgA']+rAg'\Biggr\}
-{\cal T}-a_2{\cal T}^2-4\aleph+a_0=0,
\end{eqnarray}
where ${\cal T}'\equiv d{\cal T}(r)/dr$ is calculated through (\ref{df}).

We have to note that Eq. (\ref{df31}) is a second-order algebraic equation and
 it gives
${\cal T}={\cal T}_0=const$. On the other hand, Eq. (\ref{df}) for ${\cal T}$ gives 
the solution
\begin{eqnarray}\label{df4g}
& &  A(r)= \Lambda_{eff}r^2-\frac{m}{r^{d-3}},\nonumber\\
\end{eqnarray}
where $m$ is a constant related to the mass, and the function
$g(r)$ is obtained by  (\ref{df4g}) into (\ref{df32}). Eq. (\ref{df33}) gives
 \begin{eqnarray}\label{df4gb}
&&g(r)=q(r)=c\;,\nonumber\\
&&\aleph(r)=\frac{a_1T_0+3a_2 T_0{}^2+a_0}{4}\;.
\end{eqnarray}
% \begin{eqnarray}\label{df4gb}
%h(r)=\frac{-3\beta\pm\sqrt{9\beta^2-20\gamma}}{10(N-1)(N-2)\gamma\Lambda_{eff}}=h_0.
%\end{eqnarray}
%where $c$ is another a constant of integration which cannot be set to zero since in that
%case the above solution does not exist.
In the above expressions the constant $\Lambda_{eff}$,
is given by
 \begin{eqnarray}
 \label{Leff}
\Lambda_{eff}=\frac{T_0}{3c}.
\end{eqnarray}
 It is straightforward to see that this is  an effective cosmological constant given by the torsion scalar.

The horizons of solution (\ref{df4g}) is given by
\begin{equation}\label{hor}
m=\Lambda_{eff} r^{d-1},
\end{equation}
which, in 4-dimensions, gives $m=\Lambda_{eff} r^3$.

Let us  continue our analysis for the general case
where  the torsion scalar has non-trivial values.  For a non-constant torsion scalar, we get the following solution:
\begin{eqnarray}\label{df4}
& &  A(r)=\frac{1}{r^{d-3}}\Big(\int \Big[r^{4(d-2)}q'^3c_2-\frac{a_1r^{3(d-2)}q'^2}{2a_2 (d-2)c_1}\Big]dr+c_3\Big), \nonumber\\
&&\aleph(r)=\frac{1}{16a_2}[a_1{}^2-4a_0a_2-8(d-2)r^{d-2}q' a_1a_2c_1c_2+12(d-2)^2r^{2(d-2)}q'^2 a_2{}^2 c_1{}^2c_2{}^2],\nonumber\\
&&g(r)=\frac{c_1}{q'^2r^{2(d-2)}}, \qquad q(r)=q(r),
\end{eqnarray}
where $c_i$, $i=1\cdots 3$ are integration constants.
We can assume a given  form for the arbitrary function $\aleph(r)$ and calculate the other functions from it.
So let us fix the arbitrary function to have the form
\begin{equation}\label{hor11}
\aleph(r)=-\frac{P sech^2(\frac{q_1}{(d-3)mr^{d-3}})}{r^{2(d-2)}}.
\end{equation}
It is worth noticing   that, in case $d=4$, Eq. (\ref{hor11}) is identical to  that given in \cite{1999PhLB..464...25A}. Using Eq. (\ref{hor11}) in (\ref{df4}) we get
\begin{eqnarray}\label{df44}
& &  A(r)=-\frac{1}{r^{d-3}}\Big\{\int\frac{1}{216(d-2)^3r^{2(d-2)}a_2{}^3c_1{}^3c_2{}^2(1+e^{\frac{2q_1}{(d-3)mr^{d-3}}})^3}\Big[2a_1r^{d-2}[
e^{\frac{2q_1}{(d-3)mr^{d-3}}}+1]\nonumber\\
&&-\sqrt{\mathfrak{N}r^{2(d-2)}e^{\frac{4q_1}{(d-3)mr^{d-3}}}+2[\mathfrak{N}r^{2(d-2)}-96Pa_2]
e^{\frac{2q_1}{(d-3)mr^{d-3}}}
+\mathfrak{N}r^{2(d-2)}}\Big]^2\Big[a_1r^{d-2}[
e^{\frac{2q_1}{(d-3)mr^{d-3}}}+1]\nonumber\\
&&+\sqrt{\mathfrak{N}r^{2(d-2)}e^{\frac{4q_1}{(d-3)mr^{d-3}}}+2[\mathfrak{N}r^{2(d-2)}-96Pa_2]
e^{\frac{2q_1}{(d-3)mr^{d-3}}}
+\mathfrak{N}r^{2(d-2)}}\Big]dr+c_3\Big\}, \nonumber\\
&&g(r)=\frac{36(d-2)^2a_2{}^2 r^{2(d-2)}c_1{}^3c_2{}^2(1+e^{\frac{2q_1}{(d-3)mr^{d-3}}})^2}{[2a_1r^{(d-2)}[e^{\frac{2q_1}{(d-3)mr^{d-3}}}+1]-\sqrt{\mathfrak{N}r^{2(d-2)}e^{\frac{4q_1}{(d-3)mr^{d-3}}}+
2[\mathfrak{N}r^{2(d-2)}-96Pa_2]e^{\frac{2q_1}{(d-3)mr^{d-3}}}
+\mathfrak{N}r^{2(d-2)}}]^2},\nonumber\\
&& q(r)=\int\frac{2a_1r^{d-2}[e^{\frac{2q_1}{(d-3)mr^{d-3}}}+1]-\sqrt{\mathfrak{N}r^{2(d-2)}e^{\frac{4q_1}{(d-3)mr^{d-3}}}+2[\mathfrak{N}r^{2(d-2)}-96Pa_2]
e^{\frac{2q_1}{(d-3)mr^{d-3}}}
+\mathfrak{N}r^{2(d-2)}}}{6(d-2) r^{2(d-2)}a_2c_1c_2(1+e^{\frac{2q_1}{(d-3)mr^{d-3}}})}dr+c_3,\nonumber\\
&&
\end{eqnarray}
where $\mathfrak{N}=(a_1{}^2+12a_0a_2)$.

Now if we assume the constraint $\mathfrak{N}=0$, i.e.,  ${\displaystyle a_0=-\frac{a_1{}^2}{12a_2}}$, in Eq. (\ref{df31}) we get the same solution given by Eq. (\ref{df4}) except the arbitrary function $\aleph(r)$ which takes the  form
\begin{eqnarray}\label{df444}
& &  \aleph(r)=\frac{(a_1-3(d-2)r^{(d-2)}q'a_2 c_1c_2)^2}{12a_2}.
\end{eqnarray}
Considering the function    (\ref{hor11}) in (\ref{df444}), we get
\begin{eqnarray}\label{df55}
&&A(r)=\Lambda_{eff} r^2-\frac{4(d-3)^2mq^2P}{(d-2)a_1{}^2c_1r^{d-2}q_1{}^2[e^{\frac{2q_1}{(d-3)mr^{d-3}}}+1]^2}\Big(2\sqrt{3P|a_2|}e^{\frac{q_1}{(d-3)mr^{d-3}}} [q_1\{e^{\frac{2q_1}{(d-3)mr^{d-3}}}-1\}+[1+e^{\frac{2q_1}{(d-3)mr^{d-3}}}](d-2)mr^{d-3})]\nonumber\\
&&+
3a_1q_1r^{d-2}[e^{\frac{2q_1}{(d-3)mr^{d-3}}}+1]\Big) +\frac{q^2}{(d-2)a_1{}^2c_1}\int\frac{8(d-3)^2P\sqrt{3P|a_2|}(q_1{}^2-(d-2)m^2r^{2(d-3)})e^{\frac{q_1}{(d-3)mr^{d-3}}} }{q_1{}^2r^{2(d-2)}[e^{\frac{2q_1}{(d-3)mr^{d-3}}}+1]}dr+\frac{c_3}{r^{d-3}},\nonumber\\
&&\approx\Lambda_{eff} r^2-\frac{M}{r^{d-3}}+\frac{Q^2}{r^{2(d-3)}}+\frac{Q_1{}^4}{r^{(3d-8)}}+\frac{Q_2{}^4}{r^{4(d-3)}}+\cdots, \nonumber\\
&&g(r)=\frac{a_1{}^2c_1r^{2(d-2)}[e^{\frac{2q_1}{(d-3)mr^{d-3}}}+1]^2}{q^2(d-3)^2\{a_1r^{d-2}[e^{\frac{2q_1}{(d-3)mr^{d-3}}}+1]-4\sqrt{3P|a_2|}e^{\frac{q_1}{(d-3)mr^{d-3}}}\}^2}\nonumber\\
&&\approx\frac{c_1}{(d-3)^2 q^2}\Big[1+\frac{4(d-3)^2\sqrt{3P|a_2|}}{a_1r^{d-2}}-\frac{2q_1{}^2\sqrt{3P|a_2|}}{a_1(d-3)^2 m^2r^{3d-8}}-\frac{36Pa_2}{a_1{}^2r^{2(d-2)}}+\cdots,\\
&& q(r)=\int\frac{a_1r^{d-2}[e^{\frac{2q_1}{(d-3)mr^{d-3}}}+1]-4e^{\frac{q_1}{(d-3)mr^{d-3}}}\sqrt{3P|a_2|}}{3(d-2) r^{2(d-2)}a_2c_1c_2(1+e^{\frac{2q_1}{(d-3)mr^{d-3}}})}dr+c_3\nonumber\\
&&\approx c_3+\frac{q}{r^{d-3}}- \frac{2(d-3)\sqrt{3|a_2|P}q}{a_1(2d-5)r^{2d-5}}+\frac{\sqrt{3|a_2|P}qq_1{}^2}{(d-3)(4d-11)a_1m^2r^{4d-11}}+\cdots,\nonumber
\end{eqnarray}
where $\Lambda=-\frac{a_1{}^3}{54(d-1)(d-2)^3a_2{}^3 c_1{}^3c_2{}^2}$, $q=-\frac{a_1}{3(d-2)(d-3)a_2 c_1c_2}$, $M=c_3-\frac{6(d-3)^2Pq^2}{q_1a_1c_1(d-2)}$, $Q^2=\frac{6(d-3)Pq^2}{a_1c_1(d-2)}$, $Q_1{}^4=-\frac{8(d-3)^2Pq^2\sqrt{3P|a_2|}}{c_1(d-2)(2d-5)}$ and $Q_2{}^4=-\frac{2Pq_1{}^2q^2}{m^2a_1c_1(d-2)(d-3)}$.
 Using Eq. (\ref{Max3}) we get the linear electrodynamics in the
  following form
\begin{eqnarray}\label{df4444}
& &  {\cal E}(r)=\frac{\phi}{r^{d-3}}+\frac{6\phi^2(d-2)(d-3)^2\sqrt{3|a_2|P}}{a_1r^{(2d-5)}(2d-5)}+\frac{q_1{}^2\phi^2(2d-5)}{m^2r^{3(d-3)}(d-3)}+\cdots,
\end{eqnarray}
 where $\phi=-\frac{1}{3(d-2)(d-3)}$. To get this result,  we have put $2a_2 c_1{}^2c_2=a_1{\cal P}$. It is  interesting to note that Eq. (\ref{df4444}) coincides with that given in  \cite{2017JHEP...07..136A} for   $q_1=0$. The parameter $q_1$ is  responsible  for deviations from linear electrodynamics as Eq.  (\ref{hor11}) indicates. It is straightforward to show that, from  (\ref{hor11})  for  $q_1=0$,  we return to Maxwell electrodynamics and,  for $q_1\neq 0$,  we have  non-linear electrodynamics. Explicitly,  the effect of  parameter $q_1$ appears in Eq. (\ref{df4444})   showing that the gauge potential is different  from the  one presented in \cite{2017JHEP...07..136A}.

 If we calculate the invariants of the black hole solution (\ref{df55}),  we get the same asymptotic behavior presented in \cite{2017JHEP...07..136A, Nashed:2018cth}. These invariants show that there is a  singularity at $r=0$. Approaching to $r=0$,   these invariants assume the form $(K,R_{\mu
\nu}R^{\mu \nu}) \sim \sqrt[3]{r^{-4(d-2)}}$, and  $(R,{\cal T})\sim  \sqrt[3]{r^{-2(d-2)}}$, differently  to the  black holes of Maxwell electrodynamics   in
either GR or  TEGR theories which have the forms $(K ,R_{\mu \nu}R^{\mu \nu})\sim r^{-2d}$ and $(R,{\cal T}) \sim  r^{-d}$, respectively. The above results indicate in a  clear way  that the singularity of the non-linear  charged black hole is milder than the one emerging  in GR and TEGR for
the charged case.

Finally, if we calculate the energy of solution (\ref{df55}) we get the same formula presented in \cite{2017JHEP...07..136A, Nashed:2018cth} up to the leading order, i.e. $E=\frac{(d-2)M}{4(d-3)G_d}$.In other words, this feature assures the consistency of the solution.

 \section{Rotating black holes in Maxwell-$f(T)$ gravity }\label{S5}

Let us   derive now  rotating black hole solutions  satisfying the field
equations of the above  (\ref{powellaw})  $f({\cal T})$ gravity. We start  assuming the above static
solution  as a constraint. Taking into account the following
 transformations :
\begin{equation} \label{t1}
\bar{\eta}_{i} =-\Xi~ {\eta_{i}}+\frac{ n_i}{l^2}~t,\qquad \qquad \qquad
\bar{t}=
\Xi~ t-\sum\limits_{i=1}^{\omega}n_i~ \eta_i,
\end{equation}
with  $n_i$  are    rotation parameters (their number is
$\omega= \lfloor(d - 1)/2\rfloor$ where  $\lfloor ... \rfloor$ marks the
integer part), and where we can define a parameter $\lambda$  connected to the 
$\Lambda_{eff}$ of the static solution
through
\begin{eqnarray}
\lambda
=-\frac{(d-2)(d-1)}{2 \Lambda_{eff}}.
\end{eqnarray}
Additionally,
$\Xi$ is defined as
\begin{equation}\Xi:=\sqrt{1-\sum\limits_{j=1}^{{\omega}}\frac{n_j{}^2}{\lambda^2}}.\end{equation}
Adopting the transformations (\ref{t1})
to the  (\ref{tetrad}), we obtain
\begin{eqnarray}\label{tetrad1}
\nonumber \left({h^{i}}_{\mu}\right)=\left(
  \begin{array}{cccccccccccccc}
    \Xi\sqrt{A(r)} & 0 &  -n_1\sqrt{A(r)}&-n_2\sqrt{A(r)}\cdots &
-n_{\omega}\sqrt{A(r)}&0&0&\cdots&0
\\[5pt]
    0&\frac{1}{\sqrt{A(r)g(r)}} &0 &0\cdots &0&0&0&\cdots & 0\\[5pt]
          \frac{n_1r}{\lambda^2} &0 &-\Xi r&0  \cdots &0&0&0&\cdots & 0\\[5pt]
        \frac{n_2r}{\lambda^2} &0 &0  &-\Xi r\cdots & 0&0&0&\cdots & 0\\[5pt]
        \vdots & \vdots  &\vdots&\vdots&\vdots &\vdots&\vdots& \cdots & \vdots \\[5pt]
  \frac{ n_\omega r}{\lambda^2}  &  0 &0&0 \cdots & -\Xi r&0&0&\cdots & 0 \\[5pt]
   0 &  0  &0&0 \cdots &0&r&0&\cdots & 0\\[5pt]
     0 &  0  &0&0 \cdots &0&0&r&\cdots & 0\\[5pt]
       0 &  0 &0&0 \cdots &0&0&0&\cdots & r\\
  \end{array}
\right),&\\
\end{eqnarray}
where $A(r)$ and $g(r)$ are given in 
 (\ref{df55}). Hence, for the electromagnetic potential
(\ref{df44}) we get
\begin{equation}
\label{Rotpot}
\bar{q}(r)=-q(r)\left[\sum\limits_{j=1}^{\omega} n_j d\bar{\eta}_j-\Xi
d\bar{t}\right].
\end{equation}

We have to note here that  transformation (\ref{t1}) does not alter local spacetime properties, however it changes  global properties (see  \cite{Lemos:1994xp}). This feature comes out from the fact that  it mixes compact and noncompact coordinates. As a consequence,
 vielbeins (\ref{tetrad}) and (\ref{tetrad1}) can be locally transformed into each
other but this property does not hold  globally \cite{Lemos:1994xp,Awad:2002cz}.

According to  the vielbein (\ref{tetrad1}), the metric  can be  written as
\begin{eqnarray}
\label{m1}
    ds^2=-A(r)\left[\Xi d{\bar {t}}  -\sum\limits_{i=1}^{\omega}  n_{i}d{\bar {\phi}}
\right]^2+\frac{dr^2}{A(
r)g(r)}+\frac{r^2}{\lambda^4}\sum\limits_{i=1 }^{\omega}\left[n_{i}d{\bar {t}}-\Xi \lambda^2 d{\bar{\phi}}_i\right]^2+
r^2 d\xi_k^2+\frac{r^2}{\lambda^2}\sum\limits_{i<j
}^{\omega}\left(n_{i}d{\bar {\phi}}_j-n_{j}d{\bar {\phi}}_i\right)^2,
\end{eqnarray}
where $0\leq r< \infty$, $-\infty < t < \infty$, $0 \leq \eta_{i}< 2\pi$, $i=1,2 \cdots
\omega$ and $-\infty < \xi_k < \infty$. Here   $d \xi_k^2$ is
the Euclidean metric on $(d-\omega-2)$ dimensions and $k = 1,2\cdots d-3$. It is worth  mentioning that
the static configuration (\ref{m2}) is
recovered as a particular case of the above general metric as soon as  the rotation parameters
$n_j$ are  going to zero. Furthermore, it is worth stressing   that the line-element (\ref{m1}) is derived  when the Minkowski metric  (\ref{met1})
is written  in cylindrical  coordinates, that is
\begin{eqnarray}
\label{min}
\nonumber \left(\eta_{ij}\right)=\left(
  \begin{array}{cccccccccccccc}
    -1 & 0 & 0&0&0&0&0&0&0&\cdots&0 \\[5pt]
    0&1 &0 &0 &0&0&0&0&0&\cdots & 0\\[5pt]
         0 &0 &1+\frac{n_\omega{}^2}{\lambda^2\Xi^2} &-\frac{n_\omega n_1}{l^2\Xi^2}
&-\frac{n_\omega n_
2}{\lambda^2\Xi^2}&\cdots &-\frac{n_\omega n_{\omega-1}}{l^2\Xi^2}&0&0&\cdots &0\\[5pt]
        0 &0 &-\frac{n_\omega n_1}{\lambda^2\Xi^2}
&1+\frac{n_{\omega-1}{}^2}{l^2\Xi^2} & -\frac{
n_{\omega-1}n_1}{\lambda^2\Xi^2}&\cdots&-\frac{n_{\omega-1}n_{\omega-2}}{\lambda^2\Xi^2}
&0&0&\cdots & 0\\[
5pt]
        \vdots & \vdots  &\vdots&\vdots&\vdots &\vdots&\vdots& \cdots & \vdots \\[5pt]
  0 &0 &-\frac{n_\omega a_{\omega-1}}{\lambda^2\Xi^2} &-\frac{n_{\omega-1}n_{\omega-2}}{\lambda^2\Xi^2}
&-\frac{n_{
\omega-2}n_{\omega-3}}{\lambda^2\Xi^2}&\cdots &1+\frac{n_1{}^2}{\lambda^2\Xi^2}&0&0&\cdots
&0\\[5pt]
   0 &  0  &0&0  &0&\cdots&0&1&0&\cdots & 0\\[5pt]
     0 &  0  &0&0 &0&\cdots&0&0&1&\cdots & 0\\[5pt]
       0 &  0 &0&0 &0&\cdots&0&0&0&\cdots & 1\\
 \end{array}
\right).&\\
\end{eqnarray}
Here, the interesting feature is that  the  torsion components  are vanishing.  

 %%%%%%%%%%%%%%%%%%%%%%%%%%%% Section 7 %%%%%%%%%%%%%%%%%%%%%%%%%%%%%
\section{Thermodynamical stability and Phase Transitions}\label{Thermo}
%%%%%%%%%%%%%%%%%%%%%%%%%%%%%%%%%%%%%%%%%%%%%%%%%%%%%%%%%%%%%%%%%%%%
 Black hole thermodynamics  is a fundamental  subject  in physics, because it  investigates the relation between  gravitational and quantum  regimes and it is strictly related to the thorny problem of quantum gravity. In general, there are two  main approaches   to deal with  black hole  thermodynamics: The first has been  proposed by  Gibbons and Hawking \cite{Hunter:1998qe,Hawking:1998ct},   studies  the thermal properties of  Schwarzschild  solution by applying  the Euclidean continuation. The  second method   identifies the  gravitational surface and  defines the temperature of black holes  \cite{Bekenstein:1972tm,Bekenstein:1973ur,Gibbons:1977mu}.

 In this study, we are going to apply  the second approach to understand  the thermodynamics of the AdS black hole,  derived in Eq. (\ref{df55}), and   then  to study its  stability by calculating the heat capacity and  the Gibbs free energy. The black hole (\ref{df55}) is characterized by the mass, $M$, the charges (monopole, $Q$, dipole and higher order, $Q_1$ and $Q_2$) and also by a cosmological constant $\Lambda_{eff}$.
 
\begin{figure*}
\centering
\subfigure[~The function A(r) via the radial coordinate]{\label{fig:1a}\includegraphics[scale=0.35]{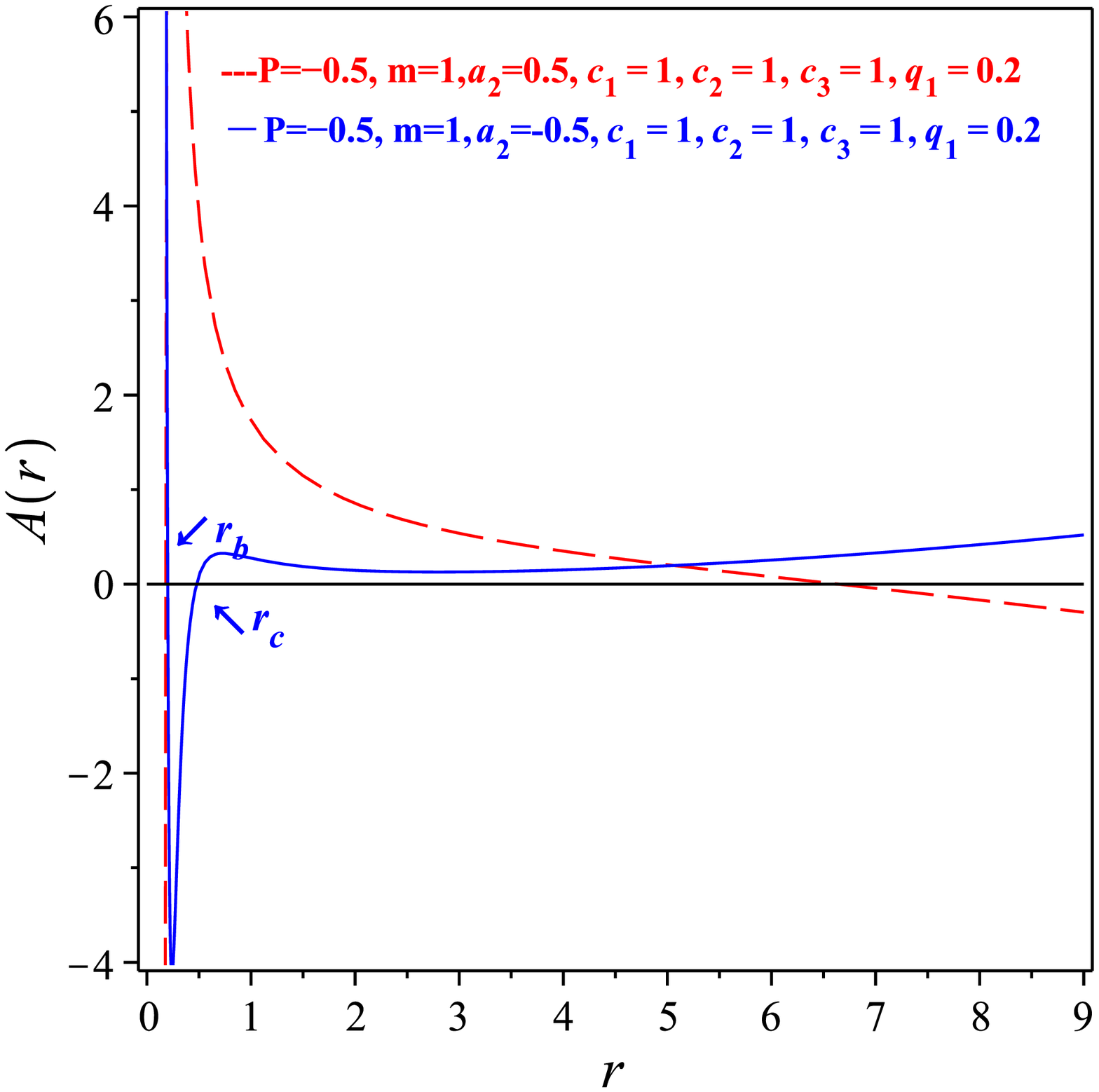}\hspace{0.2cm}}
\subfigure[~The horizon mass-radius relation]{\label{fig:1b}\includegraphics[scale=0.35]{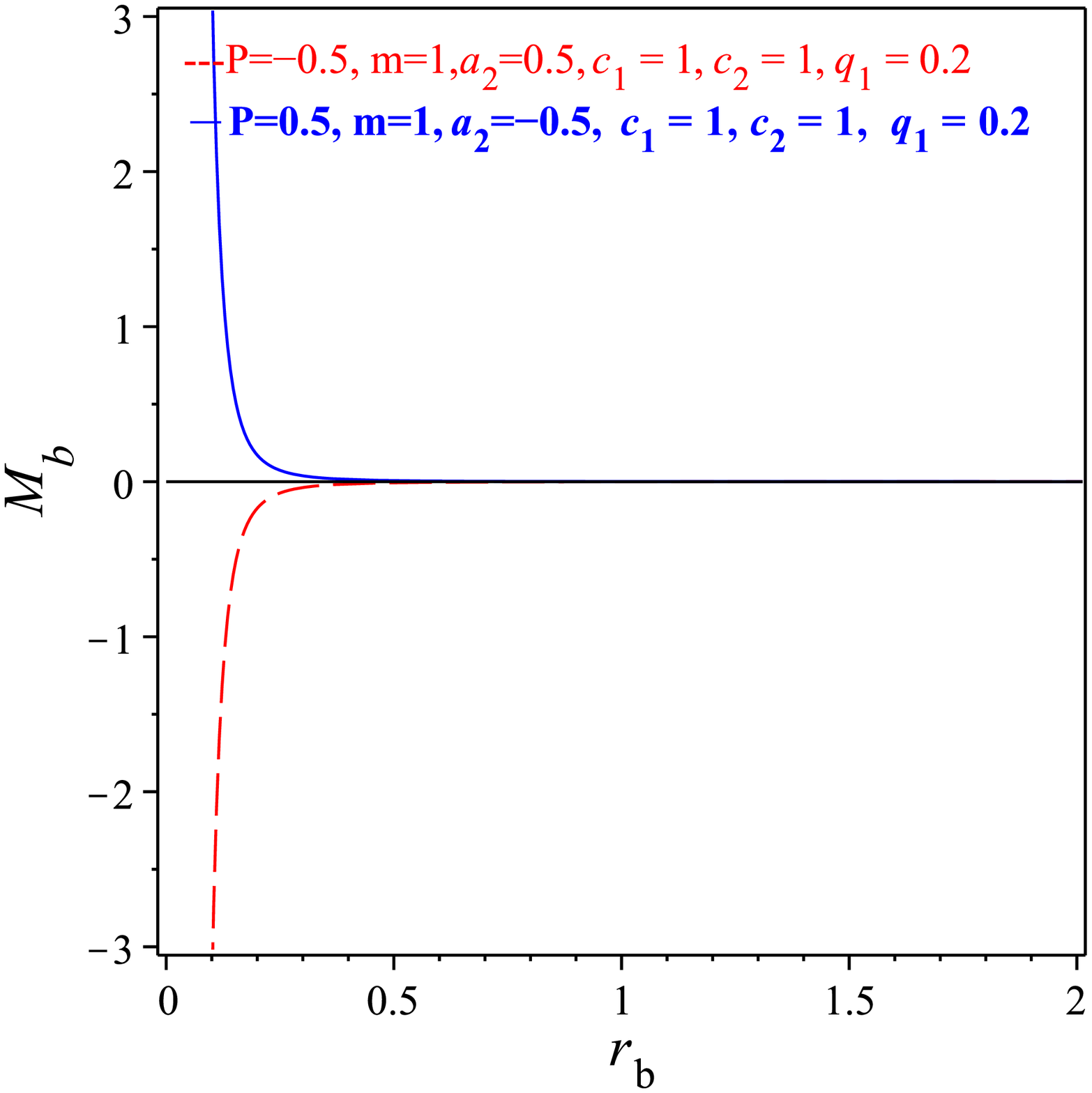}}
\caption{Schematic plot  of horizons of solution (\ref{df55}).  The plot of Fig.\ref{Fig:1} \subref{fig:1a} shows the black hole event horizon, $r_b$, and the cosmological horizon, $r_c$, while  of Fig.\ref{Fig:1} \subref{fig:1b}  shows the horizon-mass radius relation.    Here we take $d=4$ and $a_1=1$.}
\label{Fig:1}
\end{figure*}

To calculate  the horizons of  solution  (\ref{df55}), we have to put the function $A(r) = 0$. The plot of Fig.\ref{Fig:1} \subref{fig:1a}  shows the two roots of $A(r)$ which determines, respectively,  the event horizon  $r_b$  and the cosmological horizon $r_c$  of the solution  (\ref{df55}) when the dimensional parameter $a_2$ has a negative value. However, when $a_2$ has a positive  value, we have only one horizon as Fig.\ref{Fig:1} \subref{fig:1a}  shows.     In $4$-dimensions,  solutions with two horizons  can be obtained for Schwarzschild-de Sitter and Kerr-Schild black holes \cite{Di96,Dymnikova:2001fb, Dymnikova:2018uyo},  for Reissner-Nordstr\"om black holes  \cite{Ghaderi:2016dpi}, for minimal model of  regular black holes \cite{Hayward:2005gi}, and for  spherically symmetric Bardeen black holes   of  non-commutative geometry \cite{Kim:2008hm,Myung:2007av, Nicolini:2005vd,Sharif:2011ja}.
\begin{figure*}
\centering
\subfigure[~Possible horizons of the solution (\ref{df55})]{\label{fig:2a}\includegraphics[scale=0.4]{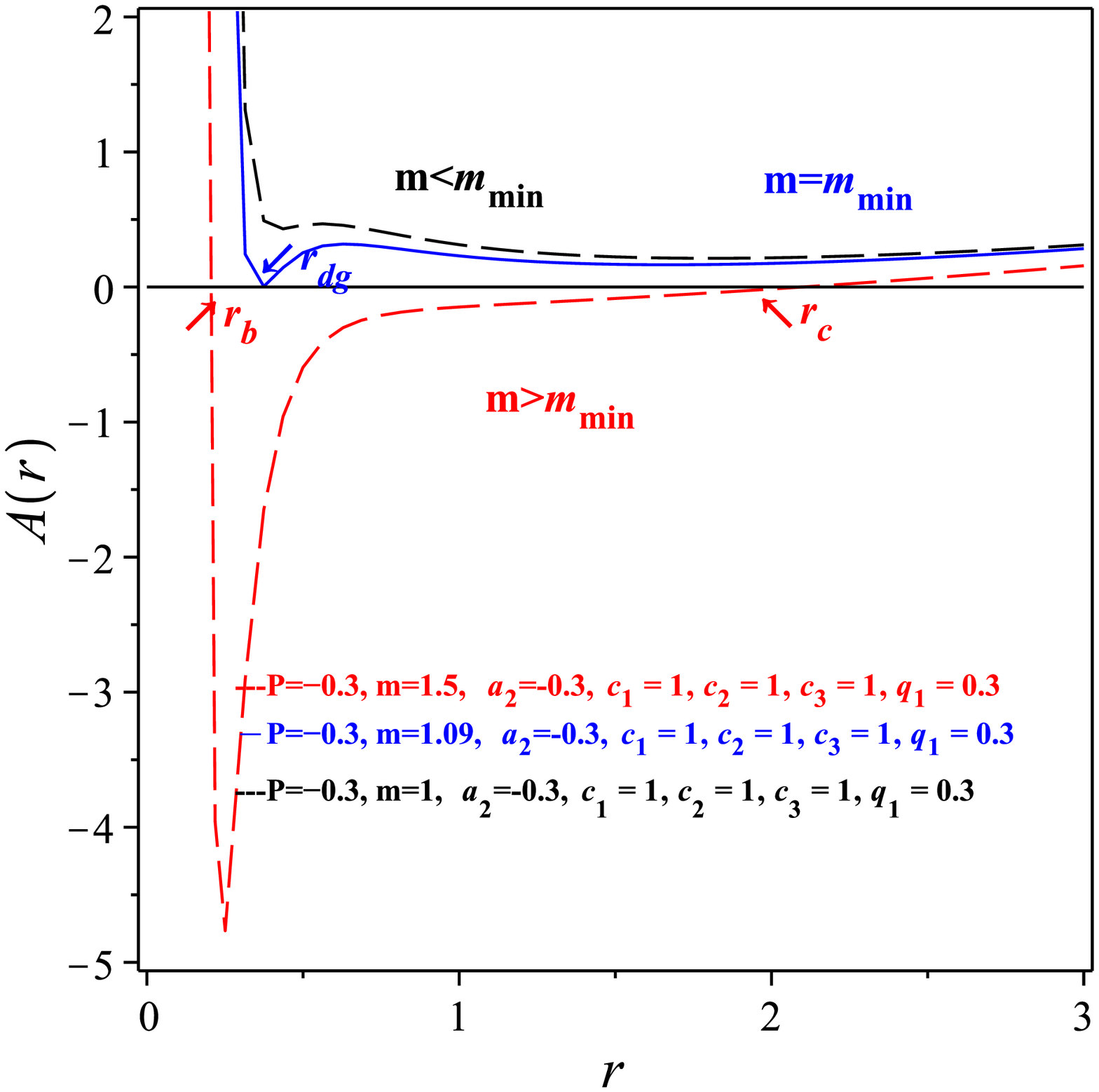}\hspace{0.2cm}}
\subfigure[~Horizon temperature--radius relation ]{\label{fig:2b}\includegraphics[scale=0.4]{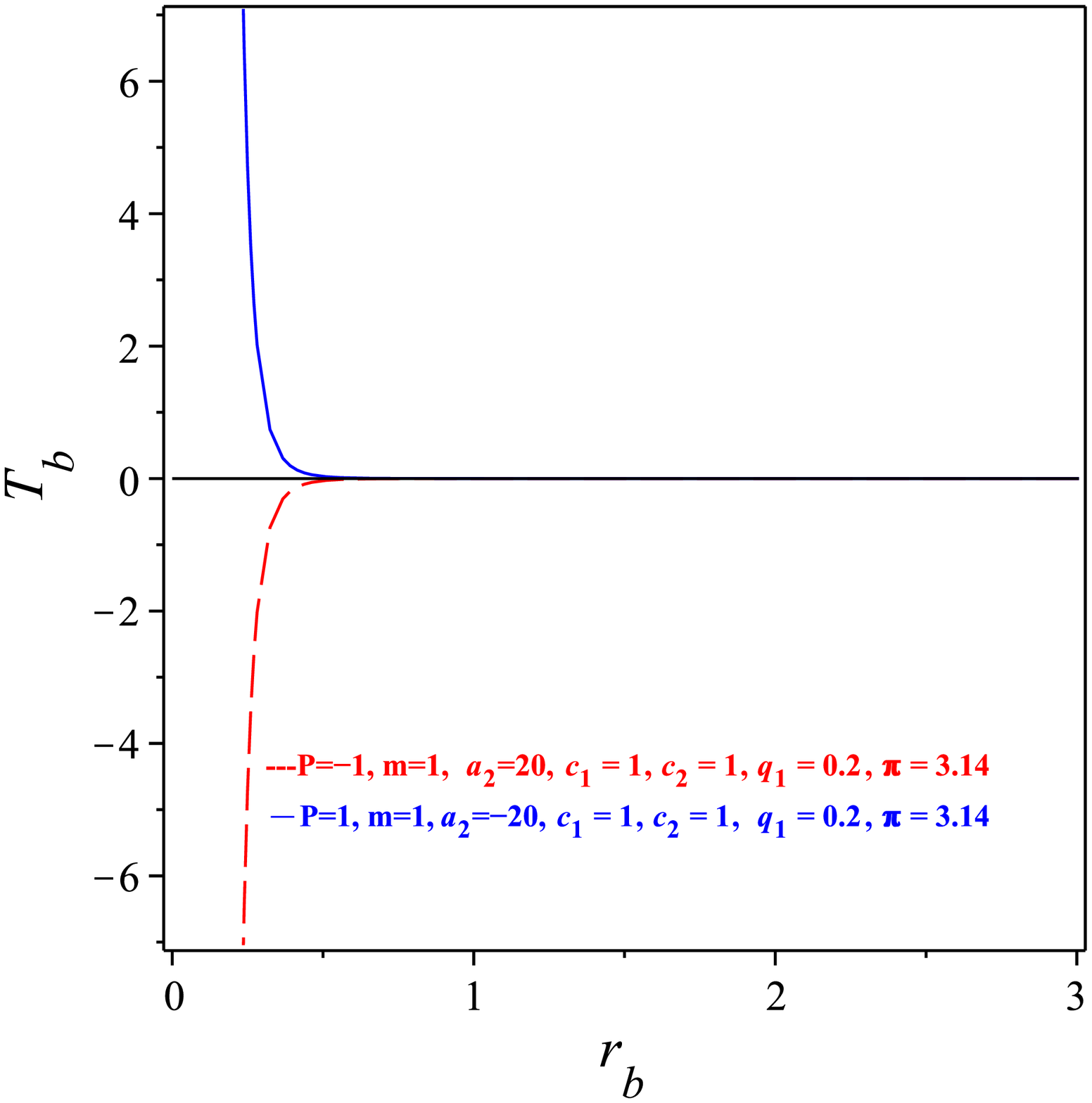}}
%\subfigure[~The degenerate horizon via $q_1$]{\label{fig:2c}\includegraphics[scale=0.25]{BF7}}
\caption{Schematic plots of the degenerate horizons of solution (\ref{df55}): \subref{fig:2a} The plot of $A(r)$ shows the black hole event horizon, $r_b$, and the cosmological horizon, $r_c$, where $m>m_{min}$. At $m=m_{min}$ ($r_b=r_c$), the black hole has a degenerate horizon at $r_{dg}$. Otherwise, $m<m_{min}$ the black hole is naked; \subref{fig:2b} the temperature vanishes on the horizon radius $r_b$.  Here we take $d=4$.}
\label{Fig:2}
\end{figure*}
\begin{figure*}
\centering
\subfigure[~Horizon heat capacity--radius relation of black hole (\ref{df55})]{\label{fig:3a}\includegraphics[scale=0.4]{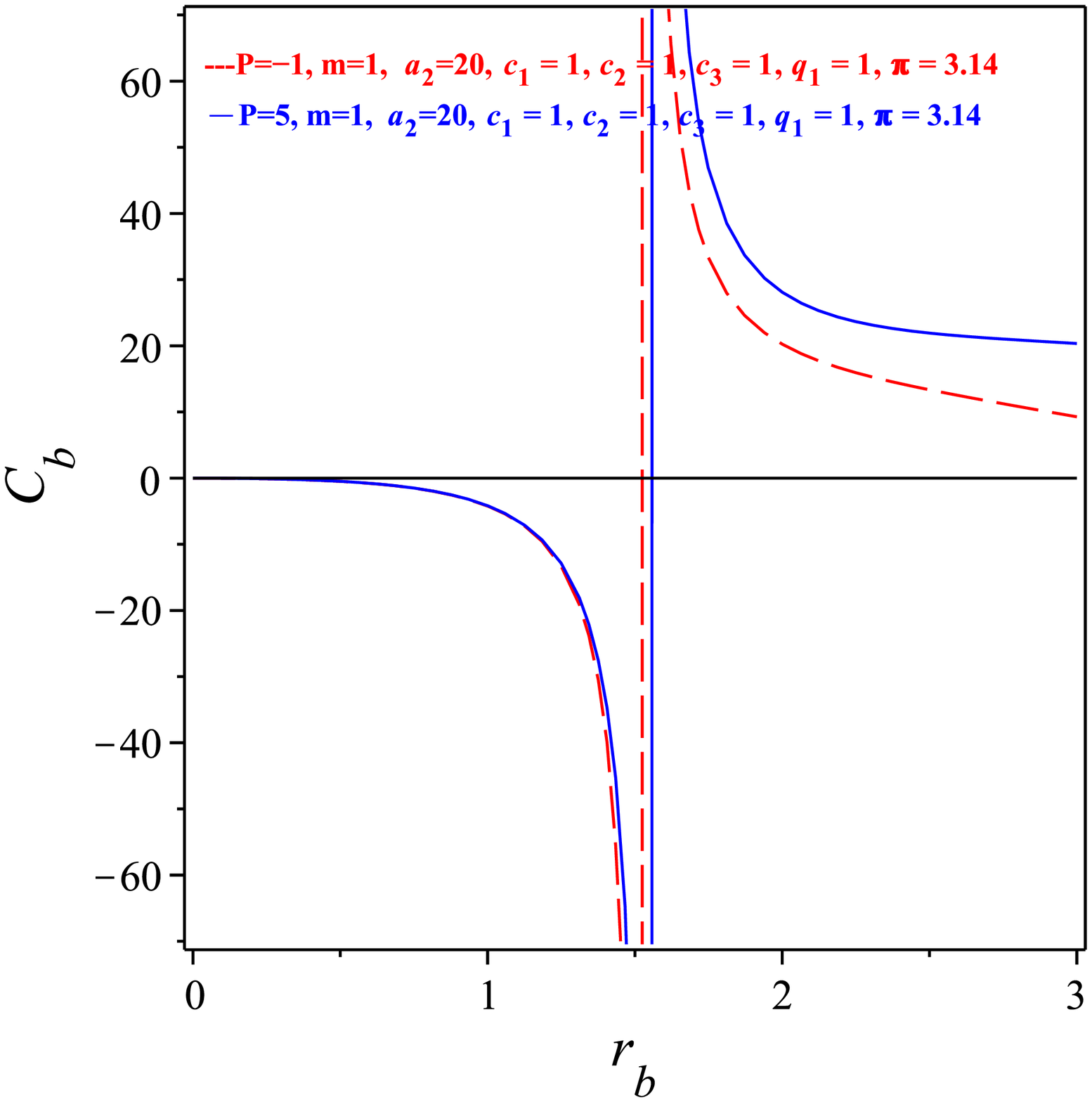}\hspace{0.2cm}}
\subfigure[~Gibbs free energy--radius relation of black hole (\ref{df55}) ]{\label{fig:3b}\includegraphics[scale=0.4]{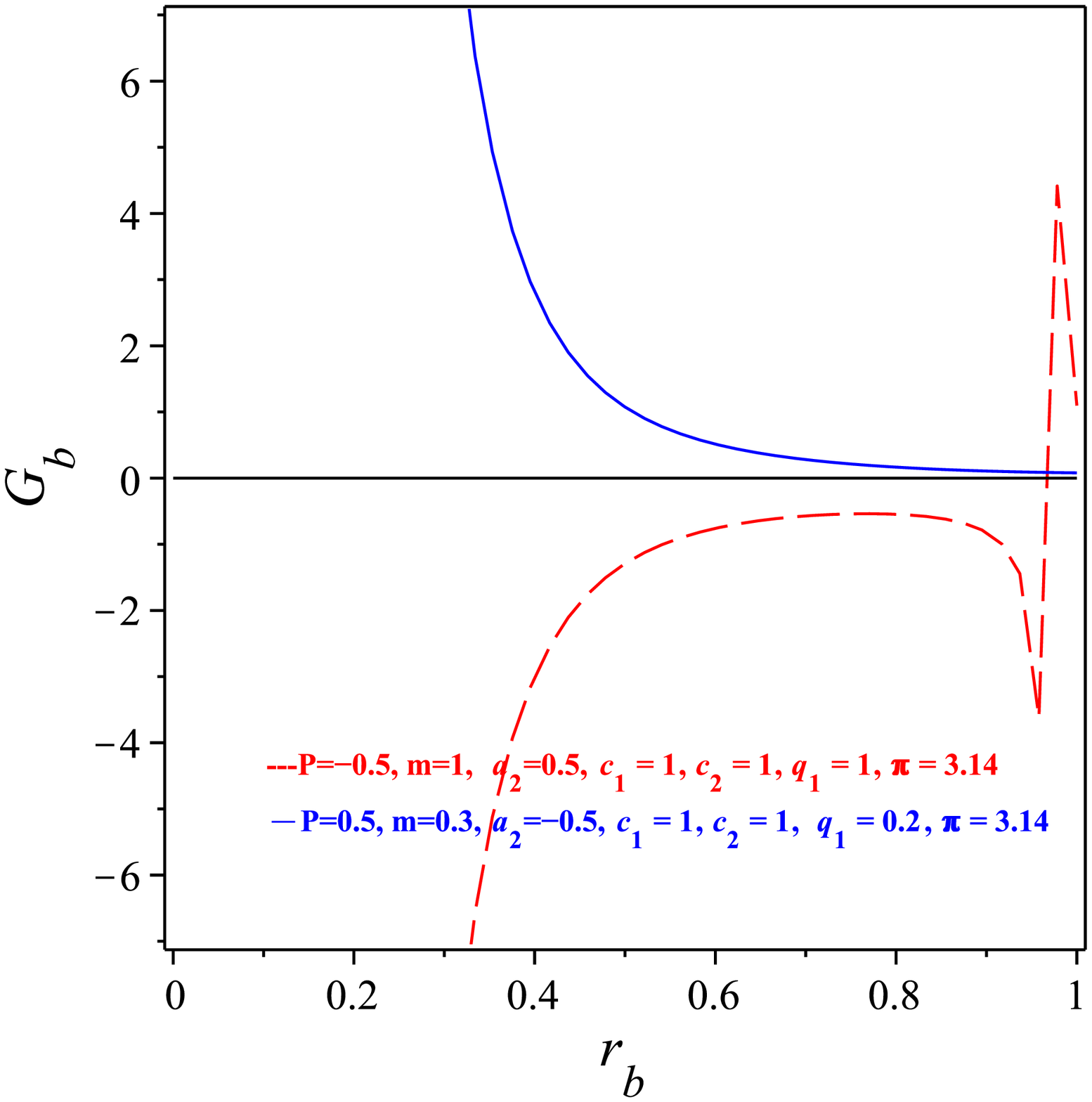}}
%\subfigure[~The degenerate horizon via $q_1$]{\label{fig:2c}\includegraphics[scale=0.25]{BF7}}
\caption{Schematic plot  of  the heat capacity shows the locally unstable event horizon which is characterized by the negative $C_h<0$. Also, the plot shows a second-order phase transition as $C_h$ diverges; \subref{fig:3b} shows free energy of black hole (\ref{df55}).}
\label{Fig:3}
\end{figure*}
%\begin{figure*}
%\centering
%\includegraphics[scale=0.35]{BF3}
%\caption{Schematic plot  of  the temperature vanishes on the horizon radius $r_b$}
%\label{Fig:3}
%\end{figure*}
%\begin{figure*}
%\centering
%\includegraphics[scale=0.35]{BF4}
%\caption{Schematic plot  of  the heat capacity shows the locally unstable event horizon which is characterized by the negative $C_h<0$. Also, the plot shows a second-order phase transition as $C_h$ diverges.}
%\label{Fig:3}
%\end{figure*}

The Bekenstein-Hawking entropy  for $f({\cal T})$ gravity can be defined as \cite{2011JCAP...11..033M}
\begin{equation}\label{ent}
S(r_b)=\frac{1}{4}Af_{\cal T}=\pi r_b{}^2f_{\cal T},
\end{equation}
where $r_b$ is the event horizon in Planck units  and $A$  is  the event horizon area. Using Eq. (\ref{df55}) in (\ref{ent}),  we get
\begin{equation}\label{ent1}
S(r_b)\approx \frac{r_b{}^{d-2}\Omega_{d-2}}{6}\left[a_1-\frac{2\sqrt{3P|a_2|}}{r_b{}^{d-2}}+\frac{q_1{}^2\sqrt{3P|a_2|}}{(d-3)^2m^2r_b{}^{3d-8}}+O\Big(\frac{1}{r_b{}^{2(d-1)}}\Big)+\cdots\right],
\end{equation}

where $\Omega_{d-2}$ is the volume of the unit $(d-2)$-sphere. Eq. (\ref{ent1}) shows that, if we neglect the higher order terms of ${\cal O}\Big(\frac{1}{r_b{}^{2(d-1)}}\Big)$,    the term   $\left(a_1-\frac{2\sqrt{3P|a_2|}}{r_b{}^{d-2}}+\frac{q_1{}^2\sqrt{3P|a_2|}}{(d-3)^2m^2r_b{}^{3d-8}}\right)$ must be positive in order to have a positive entropy. This leads to $q_1\geq \frac{\pm mr_h{}^{d-3}\sqrt{x(2x-a_1r_h{}^{d-2})}}{x}$ and $x>\frac{a_1r_b{}^{d-2}}{2}$ where $x=\sqrt{3P|a_2|}$, otherwise we have a negative entropy.

The  thermodynamical stability is related to the  heat capacity  $C_b$. In particular wth the sign of this quantity. Below, we will take into account  the thermal stability of the black holes via  their heat capacity \cite{Nouicer:2007pu,DK11,Chamblin:1999tk}
\begin{equation}\label{m55}
C_b=\frac{dE_b}{dT_b}= \frac{\partial m}{\partial r_b} \left(\frac{\partial T}{\partial r_b}\right)^{-1},
\end{equation}
where $E_b$ is the energy.
If $C_{b} > 0$ ($C_b < 0$),  the black hole is stable (unstable) from thermodynamical point of view. To  better understand this phenomenon, let us assume that,  due to thermal fluctuations, the black hole absorbs more radiation than it emits. When this happens,  its heat capacity is positive. According to this situation,   the black hole mass  increases. On the other hand, if the black hole emits more radiation than it absorbs,  the heat capacity becomes negative. In this situation,  the black hole mass  decreases and it can completely evaporate. In conclusion, black holes with negative heat capacities are  unstable from a thermodynamical point of view.

In order to calculate Eq. (\ref{m55}), we have to derive the formulae of $M_b\equiv M(r_b)$ and $T_b \equiv T(r_b)$. Firstly, we calculate the black hole mass within an even horizon $r_b$. We set $A(r_b) = 0$, then we obtain
\begin{eqnarray} \label{m33}
&&{M_b}_{{}_{{}_{{}_{{}_{\tiny Eq. (\ref{df55})}}}}}=r_b{}^{d-3}\left(\Lambda_{eff} r_b{}^2+\frac{Q^2}{r_b{}^{2(d-3)}}+\frac{Q_1{}^4}{r_b{}^{(3d-8)}}+\frac{Q_2{}^4}{r_b{}^{4(d-3)}}+\cdots\right).
\end{eqnarray}
The above equation shows that the  total mass of the black hole is given by  a function of the charge and the horizon radius. It is straightforward to calculate the degenerate horizon by the condition  $\partial M_h/\partial r_b=0$, which gives
$$r_{dg}=\pm\frac{(6a_2P)^{1/6}\sqrt{(3q_1{}^2a_1+4m^2\sqrt{3P|a_2|}+q_1\sqrt{a_1}\sqrt{9a_1q_1{}^2+24m^2\sqrt{3P|a_2|}})^{3/2}-2\sqrt{6}(m^4Pa_2)^{1/3}}}{
\sqrt{a_1}m^{1/3}(3a_1q_1{}^2+4m^2\sqrt{3P|a_2|}+q_1\sqrt{a1}\sqrt{9a_1q_1{}^2+24m^2\sqrt{3P|a_2|}})^{1/6}}, $$
 in $4$ dimensions.  As seen from Eq. (\ref{m33}) and Fig.\ref{Fig:2} \subref{fig:2a}, the horizon mass--radius relation is given by
\begin{equation} \label{m333}
M(r_b\rightarrow 0)\rightarrow \infty, \qquad \qquad M(r_b\rightarrow \infty)\rightarrow \infty.\end{equation}
%Also, one can find that there is a minimal mass at the degenerate horizon whereas the double horizons (even and cosmological) coincide. For larger masses, the double horizons are separated, while smaller masses do not show horizons.   In this sense, we find that the model at hand shares some features with the minimal model of a regular black hole \cite{Hayward:2005gi}. However, as shown by Eq. (\ref{m333}), there is no minimal length of the black hole event horizon as in the minimal model scenario.

The Hawking temperature of  black holes is derived by requiring no  singularity at the horizon of the Euclidean sector of  solutions. Furthermore, it is possible to obtain  the temperature related with the outer event horizon $r = r_b$ as \cite{Hawking:1974sw}
\begin{equation}
T = \frac{\kappa}{2\pi}, \qquad  \qquad \kappa= \frac{A'(r_b)}{2}.
\end{equation}
where $ \kappa$ is the surface gravity.
The Hawking temperature associated with the black hole solution (\ref{df55}) is
\begin{eqnarray} \label{m44}
{T_b}_{{}_{{}_{{}_{{}_{\tiny Eq. (\ref{df55})}}}}}&=&\frac{1}{4\pi}\Bigg\{(d-1)r_b \Lambda_{eff}
  -\frac{(d-3)Q^2}{r_b^{2d-5}}-\frac{(2d-5)Q_1{}^4}{r_b{}^{(3d-7)}}+\frac{3(d-3)Q_2{}^4}{r_b{}^{(4d-11)}}+\cdots\Bigg\},\nonumber\\
\end{eqnarray}
where ${T_b}$ is calculated  at the event horizon.   In Fig.\ref{Fig:2} \subref{fig:2b}, it is  shown that the horizon temperature $T_h$ is zero at the degenerate horizon $r_b=r_{dg}$. For $r_b< r_{dg}$, the horizon temperature evolves below the absolute zero giving rise to an ultra-cold black hole. As pointed out in  \cite{Davies:1978mf},  there is no  reason from thermodynamical point of view to prevent a black hole temperature to go under  absolute zero. In this case, the black hole  would become a naked singularity.    In the range $r_b > r_{dg}$, the horizon temperature is positive.  Considering also  gravitational effects, we obtain that,  for some  high temperature $T_{max}$, the radiation  becomes unstable and the collapse starts \cite{Hawking:1982dh}. As a consequence, the  AdS solution is  stable only for $T < T_{max}$. Above $T_{max}$, only the heavy black holes reach stable configurations \cite{Hawking:1982dh}.

Let us now calculate  the heat capacity $C_bh$ horizon and  substitute Eqs. (\ref{m33}) and (\ref{m44}) into Eq. (\ref{m55}). We have
\begin{eqnarray} \label{m66}
&&{C_b}_{{}_{{}_{{}_{{}_{\tiny Eq. (\ref{df55})}}}}}\approx -4\pi r_b{}^2-\frac{288\pi (d-2)\alpha  P}{r_b{}^{d-2}}+\frac{576\pi(d-2)\alpha  P\sqrt{3P|\alpha|}}{r_b{}^{2(d-2)}}+\frac{288\pi(2d-5)\alpha q_1{}^2 P }{(d-3)^2m^2r_b{}^{(3d-8)}}.  \end{eqnarray}
In Fig.\ref{Fig:3} \subref{fig:3a}, it is shown that the heat capacity is negative  for $r_b<r_{dg}$ and  positive  for $r_b > r_{dg}$. Always considering Fig.\ref{Fig:3} \subref{fig:3a}, a characteristic of the  heat capacity is a second-order phase transition at $r_{c}$ whereas the heat capacity shows an infinite discontinuity.

 The  Grand Canonical Ensemble free energy, that is the  Gibbs free energy, is defined as \cite{2012PhLB..718..687K}
\begin{equation} \label{enr}
G(r_b)=M(r_b)-T(r_b)S(r_b),
\end{equation}
where $M(r_b)$, $T(r_b)$ and $S(r_b)$ are the mass, the  temperature and the entropy  of the black hole at the event horizon, respectively.   Using Eqs. (\ref{ent}), (\ref{m33}) and (\ref{m44}) in Eq. (\ref{enr}),  we get
\begin{eqnarray}  \label{m77}
&&{G_b}_{{}_{{}_{{}_{{}_{\tiny Equation (\ref{df55})}}}}}=r_b{}^{d-3}\left(\Lambda_{eff} r_b{}^2+\frac{Q^2}{r_b{}^{2(d-3)}}+\frac{Q_1{}^4}{r_b{}^{(3d-8)}}+\frac{Q_2{}^4}{r_b{}^{4(d-3)}}+\cdots\right)+
\Omega_{d-2}\Big(\frac{q^2r_b{}^{d-1}(d-3)^2}{144(c-2)c_1\alpha}-\frac{q^2r_b(d-3)^2\sqrt{3P|\alpha|}}{72(d-2)c_1\alpha}\nonumber\\
&&+\frac{P}{36(d-2)^3c_1 r_b^{d-3}}+\frac{q^2 \sqrt{3P|\alpha|}}{144m^2(d-2)r_b^{2d-7}c_1\alpha}-\frac{5(d-3)^2Pq^2 \sqrt{3P|\alpha|}}{6(d-2)r_b^{2b-5}c_1\alpha}\Big).
\end{eqnarray}

It is worth noticing that, as soon as the charge parameter $q_1 \rightarrow 0$, the Gibbs free energy   derived from Eqs. (\ref{df55})    is coincident with that  in \cite{2014Galax...2...89A}. The  Gibbs energy of our solution is represented in Fig.\ref{Fig:3} \subref{fig:3b} for some values of  model~parameters.

\section{ Discussion and conclusions }\label{S77}
In this paper, we have investigated the effect of the non-linear electrodynamics on  modified TEGR theory. To this aim, we  derived the charged non-linear electrodynamics field equations for $f({\cal T})$ gravity. They reduce to the well known form of Maxwell field equations assuming some constrains on the arbitrary functions. Applying these field equations to cylindrical  coordinates in $d$--dimensions, we got a closed system of  non-linear differential equations.  In this framework, we obtained  black hole solutions. The most interesting feature of thes black hole solutions is that they behave as AdS solutions generalizing  the black hole solutions derived in \cite{2017JHEP...07..136A}. This generalization comes from the contribution of the parameter included in the arbitrary function (\ref{hor11}). If this parameter set equal to zero we return to the black black hole presented  in \cite{2017JHEP...07..136A}. The contributions of  non-linear electrodynamics clearly emerge in the above black holes discriminating the solutions with respect to the standard  Maxwell field. Our black holes keep all the features of the black holes derived in \cite{2017JHEP...07..136A}, i.e., they shows a central singularity, that  is softer in comparison with   the standard GR and TEGR  cases. The rotating black hole solutions can be achieved by a  suitable coordinate transformation.

More   information on the black hole (\ref{df55}) is obtained by its  thermodynamical properties. The most important feature in $f({\cal T})$ gravity  is  that  entropy is not always proportional to the horizon area  \cite{2002NuPhB.628..295C,2001IJMPA..16.5085N}.  It is possible to show that, for  constraints on the parameter $q_1$  characterizing the arbitrary function of the non-linear electrodynamics, one has a positive  entropy. On the other hand, there are some regions of  parameter $q_1$ where  entropy is  negative  \cite{2002NuPhB.628..295C,2002PhRvD..66d4012N,2017PhRvD..96j4008N,2004CQGra..21.3447C}. Negative entropy is a familiar feature in gravitational theories: several   black hole solutions have negative  entropy, e.g.   charged Gauss-Bonnet AdS black holes  \cite{2002NuPhB.628..295C,2002PhRvD..66b3522N,2002PhRvD..66d4012N,2017PhRvD..96j4008N}. Our results indicate that negative entropies may be explained as a region where the parameter $q_1$ values have entered  into an un-allowed region, or into a regime where there is a phase transition.  The  gravitational entropy of non-trivial solutions in  $f({\cal T})$ gravity will be the subject of future researches.

Furthermore, the heat capacity of black hole  (\ref{df55}) has been derived and we have shown that there is a  locally unstable event horizon  characterized by  $C_b<0$. Furthermore there is  a second-order phase transition   at $r_{b}$ whereas the heat capacity is characterized by an infinite discontinuity.  Finally,  the heat capacity of our black hole   has a stable event horizon which is characterized by  a positive value, i.e., $C_b>0$ for which $r_b>r_{dg}$.  Finally, we have derived the  Gibbs free energy  showing that the black hole solution (\ref{df55}) has always  a positive value of this quantity for some constrains on the parameter  as Fig.\ref{Fig:3} \subref{fig:3b} shows. In a forthcoming study, possible astrophysical applications of these solutions will be considered.

\section*{Acknowledgments}
 SC is supported in part by the INFN sezione di Napoli, {\it iniziative specifiche QGSKY and MOONLIGHT2}. The article is also based upon work from COST action CA15117 (CANTATA), supported by COST (European Cooperation in Science and Technology).

%%%%%%%%%%%%%%%%%%%%%%%%%%%%%%%%%%%%%%%%%%%%%%%%%%%%%%%%%%%%%%%%%%%%%%%%%%%%%%%%%%%%%%
%\bibliographystyle{apsrev}
%\bibliography{Ref}

%%%%%%%%%%%%%%%%%%%%%%%%%%%%%%%%%%%%%%%%%%%%%%%%%%%%%%%%%%%%%%%%%%%%%%%%%%%%%%%%%%%%%%
\end{document}